\documentclass[aps,prb,letterpaper,amsmath,amssymb,superscriptaddress,reprint]{revtex4-1}
\usepackage{graphicx}

\begin{document}
\title{Electronic structure and optical band gap determination of NiFe$_2$O$_4$}
\author{Markus Meinert}
\email{meinert@physik.uni-bielefeld.de}

\affiliation{Center for Spinelectronic Materials and Devices, Department of Physics, Bielefeld University, D-33501 Bielefeld, Germany}
%\author{Christoph Klewe}
\author{G\"unter Reiss}
\affiliation{Center for Spinelectronic Materials and Devices, Department of Physics, Bielefeld University, D-33501 Bielefeld, Germany}
\date{\today}

\begin{abstract}
In a theoretical study we investigate the electronic structure and the band gap of the inverse spinel ferrite NiFe$_2$O$_4$. The experimental optical absorption spectrum is accurately reproduced by fitting the Tran-Blaha parameter in the modified Becke-Johnson potential. The accuracy of the commonly applied Tauc plot to find the optical gap is assessed based on the computed spectra and we find that this approach can lead to a misinterpretation of the experimental data. The minimum gap of NiFe$_2$O$_4$ is found to be a 1.53\,eV wide indirect gap, which is located in the minority spin channel.
\end{abstract}

\maketitle

Today, DFT is the main tool to obtain the electronic structure of solids.\cite{HK, KS} A long-standing problem of electronic structure theory is the description of transition metal oxides. These exhibit strong electron-electron correlation, which is not properly accounted for by the density functional theory (DFT) with approximate local functionals. Here we focus on NiFe$_2$O$_4$, a ferrimagnetic inverse spinel ferrite,\cite{Hastings53, Youssef69} which poses an example of such difficult to describe systems. Experimental investigations mostly based on optical absorption on this material found band gaps between 1.5\,eV and 5\,eV.\cite{Laan99, Joshi03, Balaji05, Dolia06, Srivastava09, Chavan10, Bharathi11, Rai12, Sun2012} Theoretical investigations on the electronic characteristics of bulk NFO using a self-interaction-corrected local spin-density approximation (SIC-LSDA) approach,\cite{Szotek06} or by including a Hubbard correction in terms of the DFT$+U$ method\cite{Antonov03,Fritsch2010} have predicted a bandgap of around 1\,eV. Sun \textit{et al.} performed band structure calculations using DFT$+U$ and a hybrid functional (HSE06).\cite{Sun2012} They obtained a bandgap of 2.7\,eV with HSE06 and 1.6\,eV for the DFT$+U$ computations. Thus, there is still a lot of controversy on the band structure and the gap of NiFe$_2$O$_4$.

The appropriate framework to discuss electron correlations and band structures is the many-body perturbation theory, e.g., within the $GW$ approximation.\cite{Aryasetiawan98, Shishkin07} Unfortunately, this approach is computationally very expensive. Tran and Blaha recently proposed an alternative, similarly accurate and computationally cheaper method to obtain the band gap directly as differences of Kohn-Sham eigenvalues: they modified the Becke-Johnson exchange potential\cite{BJ06} with a parameter $c$, so that it reads\cite{TB09}
\begin{equation}
v_{x,\sigma}^\mathrm{mBJ}(\mathbf{r}) = c v_{x,\sigma}^\mathrm{BR}(\mathbf{r}) + (3c -2) \frac{1}{\pi} \sqrt{\frac{5}{12}} \sqrt{\frac{2 t_\sigma(\mathbf{r})}{n_\sigma(\mathbf{r})}},
\end{equation}
where $n_\sigma(\mathbf{r})$ is the spin-dependent electron density and $t_\sigma(\mathbf{r})$ is the spin-dependent kinetic-energy density. $v_{x,\sigma}^\mathrm{BR}(\mathbf{r})$ is the Becke-Roussel potential, which models the Coulomb potential created by the exchange hole.\cite{BR89} Due to the kinetic-energy dependent term in the mBJ potential, it reproduces the step-structure and derivative discontinuity of the effective exact exchange potential of free atoms.\cite{Armiento08} 
The parameter $c$ was proposed to be determined self-consistently from the density and is related to the dielectric response of the system.\cite{Marques11, Krukau08} $c$ increases with the gap size and has a typical range of 1.1--1.7.\cite{TB09} The mBJ potential has been proposed to be combined with LDA correlation (mBJ\-LDA). Its particular merits and limits have been reviewed by Koller \textit{et al.}\cite{Koller11} % In a recent paper, Koller \textit{et al.} have suggested a new and more balanced parametrization of $c$, based on a larger test set of solids.\cite{Koller12} It gives overall better band gaps and reduces the general underestimation of band gaps of the original parametrization.

In recent publications, the performance of mBJ\-LDA for complete band structure calculations rather than just band gap predictions has been discussed. For simple semiconductors it was found that the band widths are too small if $c$ is adjusted to get the correct band gap.\cite{Kim10} It is also unsuitable to describe half-metals.\cite{Meinert13} However, mBJ\-LDA predicts the unoccupied band structure of NiO and optical spectra of TiO$_2$ with good accuracy.\cite{Hetaba12, Gong12} In this communication, we compare optical absorption spectra of NiFe$_2$O$_4$ thin films with computational results using the mBJ\-LDA potential.

The calculations in this work are based on the full-potential linearized augmented-plane-wave (FLAPW) method and were done with the \textsc{elk} code.\cite{elk} The mBJ exchange potential is available through an interface to the \textsc{Libxc} library.\cite{Libxc} A $10 \times 10 \times 10$ $\mathbf{k}$-point mesh with 171 inequivalent points was used for the Brillouin zone integration. The muffin-tin radii were set to 1.8\,bohr for the transition metals and 1.7\,bohr for O. The mBJ exchange potential was coupled with the Perdew-Wang LDA correlation.\cite{PW92} 
We have used the experimental lattice constant of $a = 8.33$\,\AA{},\cite{Youssef69} and relaxed the internal atomic coordinates using the PBE functional.\cite{PBE} The dielectric function was computed in the independent particle approximation. In the inverse spinel structure, the transition metals sites surrounded by O tetrahedra are occupied with Fe, while the octahedral sites are randomly occupied with Fe and Ni. We have to use an an ordered cell instead of the proper disordered unit cell in the calculation, so the symmetry is artificially reduced from \textit{Fd$\mathit{\bar{3}}$m} to \textit{Imma}. Thus, we take the observable macroscopic dielectric function to be $\varepsilon_\mathrm{M}(\omega) = 1/3\, \mathrm{Tr}\, \varepsilon_{ij}(\omega)$ to restore the full symmetry. The spectra were broadened with an 80\,meV wide Lorentzian. The effect of excitons on the absorption spectrum was investigated with time-dependent DFT (TDDFT) using the bootstrap kernel.\cite{Sharma11}

\begin{figure}[t]
\includegraphics[width=8.6cm]{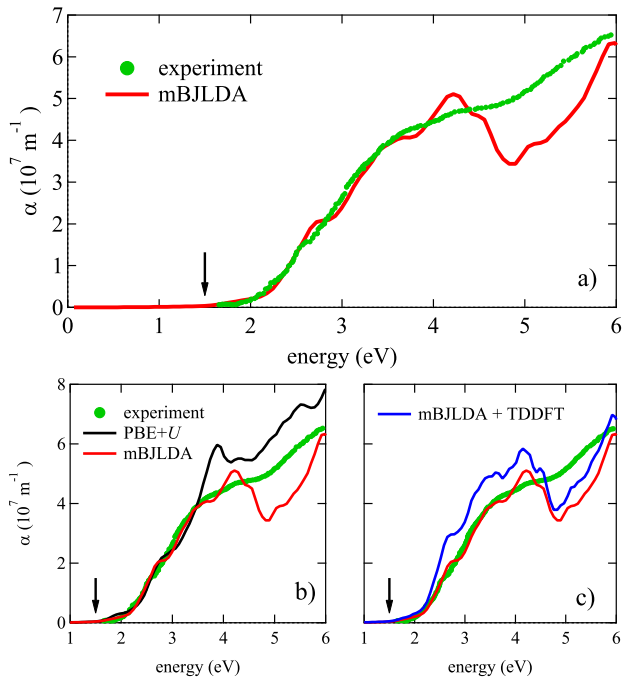}
\caption{\label{optics}a): Optical absorption spectrum of a NiFe$_2$O$_4$ thin film from Ref. \onlinecite{Holinsworth13} and mBJ\-LDA calculated absorption spectrum. b): Comparison of mBJ\-LDA and PBE$+U$ calculated absorption spectra. c): Comparison of noninteracting and bootstrap TDDFT mBJ\-LDA absorption spectra. The arrow marks the mBJ\-LDA calculated fundamental gap.}
\end{figure}

In Fig. \ref{optics} a) we compare the experimental optical absorption spectra of high-quality pulsed laser deposited thin films of NiFe$_2$O$_4$ with our mBJ\-LDA calculation.\cite{Sun2012, Ma10, Holinsworth13} The Tran-Blaha parameter $c = 1.44$ has been chosen such that the computed absorption spectrum matches the experimental spectrum between 2 and 2.5\,eV. We note that the density-based $c$ from the original Tran-Blaha paper is just slightly smaller, $c_\mathrm{TB} = 1.42$. The overall agreement between experiments and mBJ\-LDA calculation is remarkably good. The agreement is remarkable in view of the fact that an accurate absorption spectrum requires a good description of both valence and conduction states. However, some spectral weight around 5\,eV is missing in the calculation. The agreement up to 4.5\,eV is somewhat better than for a PBE$+U$ calculation, which we show in Fig. \ref{optics} b). Here, the Hubbard parameters\cite{Liechtenstein95} have been chosen as $U_\mathrm{Fe,Ni} = 4.5$\,eV and $J_\mathrm{Fe,Ni} = 0.9$\,eV. $U_\mathrm{Fe}$ governs the size of the gap; changing it leads to a rigid shift of the absorption spectrum up to 5\,eV. It was chosen to match the mBJ\-LDA gap. The choice of $U_\mathrm{Ni}$ is not critical and only leads to modifications of the absorption spectrum above 5\,eV. We attribute the good reproduction of the absorption spectrum by mBJ\-LDA to a more accurate description of the O $p$ states, which are more localized in the mBJ\-LDA calculation. While the PBE$+U$ calculation only corrects the transition metal $d$ states (with respect to a plain PBE calculation), mBJ\-LDA allows for an improved description of all electrons. This will be discussed in more detail later.

Bound excitons play no significant role for the optical properties of NiFe$_2$O$_4$ as is shown in Fig. \ref{optics} c). In the TDDFT calculation, the absorption is enhanced by 20 to 40\,\% (at odds with experiment), but the spectral features do not shift to lower energies. Due to the small band gap, the screening is strong: the ion-clamped static dielectric constant is $\varepsilon_\infty^\mathrm{mBJ\-LDA} = 5.4$. Thus, no localized Frenkel excitons are expected to show up, as is confirmed numerically by the TDDFT calculation. While the bootstrap kernel does well in describing Frenkel excitons, it fails to describe the delocalized Wannier excitons.\cite{Yang13} These, however, are typically rather weak with binding energies of less than 0.1\,eV for materials with similar gaps and dielectric constants.\cite{Dvorak13} As we will show later, due to the particular localization of conduction band minimum (CBM) and valence band maximum (VBM) states, also the binding energies for Wannier excitons are expected to be small.\cite{Dvorak13}

\begin{figure}[t]
\includegraphics[width=8.6cm]{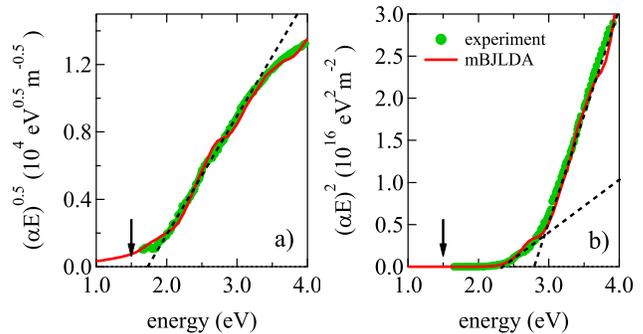}
\caption{\label{tauc}a): Tauc plot of $(\alpha E)^{0.5}$ for indirect gaps. b): Tauc plot of $(\alpha E)^2$ for direct gaps.}
\end{figure}

A common way to extract the indirect and direct gaps from optical absorption spectra is the Tauc plot, which is based on the assumption that the energy-dependent absorption coefficient $\alpha(E)$ can be expressed as\cite{Mott79, Stenzel05}
\begin{align}
\alpha (E) \, E&=  A \left( E - E_{g}^\mathrm{direct} \right)^{0.5} \nonumber \\
 &+ B \left( E - E_g^\mathrm{indir} \pm E_\mathrm{phon} \right)^2
\end{align}
with two parameters $A$ and $B$, the indirect and direct gaps $E_\mathrm{g}$, and the phonon energy $E_\mathrm{phon}$. Thus, straight line segments in $(\alpha E)^2$ indicate direct gaps and straight line segments in $(\alpha E)^{0.5}$ indicate indirect gaps. In Fig. \ref{tauc} we show that particularly the $(\alpha E)^{0.5}$-plot does not indicate an indirect gap in NiFe$_2$O$_4$:  both the experimental as well as the theoretical Tauc plot show identical straight line segments. However, the theoretical Tauc plot can by no means indicate indirect transitions, because these are not included in the calculation. Furthermore, the computed fundamental gap on which the theoretical spectrum is based is 1.53\,eV, while the Tauc plots indicate an indirect gap of about 1.65\,eV. 

\begin{figure}[t]
\includegraphics[width=8.6cm]{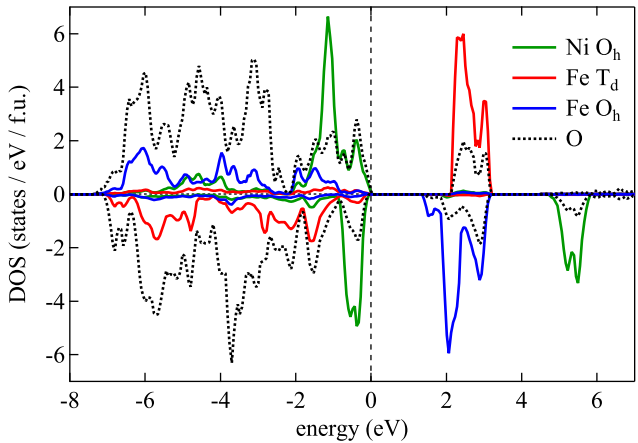}
\caption{\label{dos}Site-projected density of states of NiFe$_2$O$_4$ obtained from an optimized mBJ\-LDA calculation. Majority states are shown on positive, minority states are shown negative scale. The energy is set to zero at the valence band maximum.}
\end{figure}

Having established that mBJ\-LDA provides a good description of the electronic structure of NiFe$_2$O$_4$, we go into more detail. Table \ref{moments} summarizes the calculated magnetic spin moments and valence charges inside the muffin-tin spheres. The data for Ni are in good agreement with an ionic Ni$^{2+}$ configuration. For Fe, the valence charges are actually too large and the magnetic moments are too small for the anticipated Fe$^{3+}$ configuration.\cite{Youssef69} However, it has been shown for Fe$_3$O$_4$ that the nominal Fe$^{3+}$ species have a somewhat larger charge (lower oxidation state), which agrees with our calculation for NiFe$_2$O$_4$.\cite{Arenholz06} A substantial amount of charge ($8.84\, e^-/\mathrm{f.u.}$) is in the interstitial region between the muffin-tin spheres and accounts for the missing charge of the O$^{2-}$ ions. This number is larger in the PBE$+U$ calculation ($9.63\, e^-/\mathrm{f.u.}$), indicating the weaker localization of the O states discussed earlier. In Fig. \ref{dos} we show the site-projected density of states (DOS), which reveals that the fundamental gap is in the minority states. The VBM is mostly composed of Ni and O states and has a small exchange splitting of 0.08\,eV. This is in contrast to DFT$+U$ and HSE06 calculations, which predict a significant exchange splitting of the VBM.\cite{Sun2012} Still, the overall shape of the DOS is very similar to the HSE06 calculation. The conduction states below 6\,eV are composed of the transition metal $d$ states, which hybridize weakly with the O atoms. While the states of the two Fe species have about the same energy, the Ni states are clearly set off. This leads to the dip around 5\,eV in the computed absorption spectra which is much less pronounced in the experiment. Thus, the unoccupied Ni $d$ states are actually about 0.5\,eV lower in energy. Due to the different localization of VBM (mostly on Ni and O) and CBM (mostly on Fe and O), electron-hole pairs generated in photoabsorption are well separated, which leads to a vanishing binding energy of Wannier excitons.\cite{Dvorak13} This spatial separation and correspondingly small wavefunction overlap of the states defining the band gap is also responsible for the tiny optical absorption below 2\,eV, which makes the optical determination of the fundamental gap difficult.

\begin{table}[b]
\begin{ruledtabular}
\begin{tabular}{l c c c c c}
			&	Ni			&	Fe(T$_d$)	&	Fe(O$_h$)	&	O(1)	&	O(2) \\\hline
$m$			&	1.75			&	-3.87				&	4.08				&	0.09	&	-0.01\\
$n_\mathrm{V}$	&	7.77			&	5.25				&	5.42				&	5.68	&	5.68
\end{tabular}
\end{ruledtabular}
\caption{\label{moments} Magnetic spin moments $m$ and valence charges $n_\mathrm{V}$ inside the muffin-tin spheres as calculated by mBJ\-LDA for the transition metals and the two O types.}
\end{table}

\begin{figure}[t]
\includegraphics[width=8.6cm]{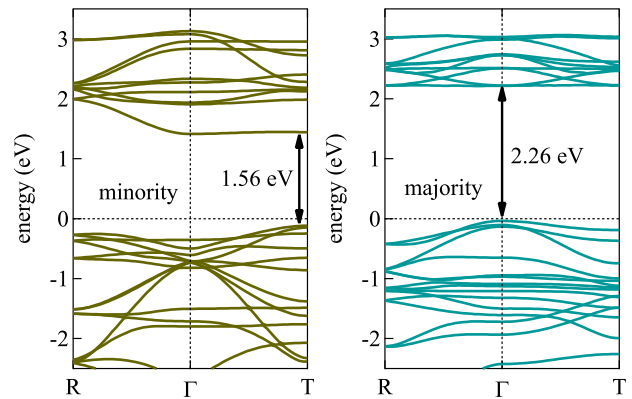}
\caption{\label{bands}Band structure plots of NiFe$_2$O$_4$ calculated with mBJ\-LDA. The high-symmetry points $R$ and $T$ correspond to the $X$ point of the disordered unit cell of the inverse spinel structure with full cubic symmetry.}
\end{figure}

Fig. \ref{bands} displays the band structure plots calculated with mBJ\-LDA. The high-symmetry points $R$ and $T$ correspond to the $X$ point of the cubic cell with full symmetry. In the real, disordered case, the dispersion along the $\Gamma - X$ path will smear out and form intermediate states defined by the $\Gamma - R$ and $\Gamma - T$ dispersions. The minority gap of NiFe$_2$O$_4$ is found to be a 1.53\,eV wide indirect gap between $T$ and $\Gamma$. However, it is only 0.03\,eV smaller than the minimum direct gap in the minority states at the $T$ point and is thus expected to play no significant role, particularly at room temperature. The minimum gap of the majority states is a 2.26\,eV wide direct gap at $\Gamma$. The Tauc plot in Fig \ref{tauc} b) indicates two direct gaps at 2.35\,eV and 2.8\,eV. The first one could correspond to the onset of majority absorption, but could equally well be due to the onset of absorption into the second unoccupied minority band. Moreover, there is no gap in the band structure that could correspond to the 2.8\,eV Tauc gap. Thus, the $(\alpha E)^2$ Tauc plot erroneously assigns a structure in the absorption spectrum to a gap, which in fact has its origin in the particular features of the band structure. Consequently, this type of plots is unsuitable to determine the band gap of NiFe$_2$O$_4$ and its use may have contributed to the broad range of experimental band gaps found in the literature.

In conclusion we have shown that the mBJLDA potential is well suited to describe the electronic structure of NiFe$_2$O$_4$. Based on the computed optical absorption spectrum we have shown that the commonly applied Tauc plot is unsuitable to determine the band gap of NiFe$_2$O$_4$ and that it can not correctly distinguish between indirect and direct transitions in this material. These findings exemplify that the Tauc plot can not in general be straightforwardly applied to materials with complex band structures and small overlap between the states which define the gap. 

We thank Christoph Klewe for inspiring discussions and the developers of the \textsc{Elk} code for their efforts. Financial support by the Deutsche Forschungsgemeinschaft under contract number RE1052/24-1 is gratefully acknowledged.

\end{document}